\documentclass[preprint]{aastex6}

\usepackage{bm}
\usepackage[first=0,last=9]{lcg}

\usepackage{color}
\definecolor{orange}{rgb}{1,0.647,0}

\begin{document}

\title{Direct estimates of the solar coronal magnetic field using contemporaneous extreme-ultraviolet, radio, and whitelight observations}

\author{Anshu Kumari\altaffilmark{1}, R. Ramesh\altaffilmark{1}, C. Kathiravan\altaffilmark{1}, T. J. Wang\altaffilmark{2,3} and N. Gopalswamy\altaffilmark{3}}
\altaffiltext{1}{Indian Institute of Astrophysics, 100 Feet Road, Koramangala II Block, Bangalore, Karnataka, India - 560034}
\altaffiltext{2}{Department of Physics, The Catholic University of America, Washington, DC 20064, USA}
\altaffiltext{3}{NASA Goddard Space Flight Center\, Code 671, Greenbelt, MD 20771, USA}
\begin{abstract}

We report a solar coronal split-band type II radio burst that was observed 
on 2016 March 16 with the Gauribidanur Radio Spectro-Polarimeter (GRASP) in the frequency 
range $\approx$\,90\,-\,50 MHz, and the Gauribidanur RadioheliograPH (GRAPH) at two discrete frequencies, viz. 80 MHz and 53.3 MHz. Observations around the same epoch in extreme-ultraviolet (EUV) and white-light show that the above burst was associated with a flux rope structure and a coronal mass ejection (CME), respectively.
The combined height-time plot generated using EUV, radio, and whitelight data suggest that the different observed features (i.e. the flux rope, type II burst and the CME) are all closely associated.
We constructed an empirical model for the coronal electron density distribution
($N_{e}(r)$, where $r$ is the heliocentric distance) from the above set of observations themselves and used it to estimate the coronal magnetic field
strength ($B$) over the range of $r$ values in which the respective events were observed.  The $B$ values are consistent with each other. They vary as $B(r)\,=\,2.61 \times r^{-2.21}$ \textrm{G} in the range $r \approx$\,1.1\,-\,2.2$\rm R_{\odot}$. As far as we know, similar `direct' estimates of $B$ in the near-Sun corona without assuming a model for $N_{e}(r)$, and by combining co-temporal set of observations in
two different regions (radio and whitelight) of the electromagnetic spectrum, have rarely been reported. Further, the present work is a novel attempt where the characteristics of a propagating EUV flux rope structure, considered to be the signature of a CME close the Sun, have been used to estimate $B(r)$ in the corresponding distance range.
\end{abstract}


\section{Introduction} \label{sec:intro}

The formation, evolution, and characteristics of coronal mass ejections, coronal streamers, coronal holes, and coronal loops in the solar atmosphere are primarily determined by the coronal magnetic field. 
But measurements of the solar coronal magnetic field
are presently limited due to practical difficulties (see for e.g., \cite{Lin2000,Tomczyk2008}). It is inferred by extrapolating the observed solar surface magnetic field distribution using the potential or force-free field approximations (see for e.g. \cite{Wiegelmann2017} for a recent review on the subject). 
Estimates of the coronal magnetic field strengths, particularly in the `middle' corona ($r\approx1.1\,-\,3.0\rm R_{\odot}$)
are largely obtained using observations of either the circularly polarized radio emission (i.e.\, the Stokes V emission) from the transient low frequency ($\lesssim$ 150 MHz) radio events like the type I, II, III, IV, and V bursts, or the split-band feature exhibited by some of the radio type II bursts \citep{Smerd1975,Dulk1978,Dulk1980,Gopalswamy1986,Bastian2001,Vrsnak2002,Mancuso2003,Mancuso2013_1,
Ramesh2003,Ramesh2004,Ramesh2011b,Ramesh2013,Cho2007,Zimovets2012,
Sasikumar2013b,Sasikumar2014,Tun2013,Hariharan2014,Zucca2014,Kishore2016,Kishore2017}. 
Weak circularly polarized component in the thermal radio emission from discrete sources at low frequencies \citep{Sastry2009,Ramesh2010a}, and geometrical properties of the propagating disturbances observed in EUV images of the solar atmosphere \citep {Gopalswamy2012} have also been used to estimate coronal magnetic strength. \cite{Kwon2013b} carried out global coronal seismology from the propagation speed of a fast magtenosonic wave to determine  $B(r)$ in the extended corona.
Despite all these different measurements, a combined estimate of $B(r)$ using observations in the different regions of the electromagnetic spectrum and particularly close to the Sun
are very limited \citep{Dulk1978,Vrsnak2002,Mancuso2003,Mancuso2019,Cho2007,Zimovets2012,Zucca2014,
Anshu2017b,Anshu2017c}.
Equally rare are reports where the same set of observations are used 
to independently derive the coronal electron density ($N_{e}(r)$) required to estimate $B(r)$. 
This is important since $B(r)$ will be otherwise sensitive to the  
density model used (see for e.g., \cite{Vrsnak2002}).

In the present work we take advantage of the simultaneous imaging and spectro-polarimetric observations of a type II radio burst with the ground based facilities, and EUV, whitelight observations of the solar corona with instruments onboard space platforms, 
to estimate $B({r})$ in the distance range $r \approx$\,1.1\,-\,2.2$\rm R_{\odot}$. The paper is arranged as follows: in Section 2, we have reported the observations and the related instruments. The data analysis and results are discussed in Section 3 with summary in Section 4.

\section{Observations} \label{sec:obs}

\subsection{Radio Observations} \label{sec:rad}

\begin{figure}[t!]
\centering
\centerline{\includegraphics[angle=0,height=10cm,width=10cm]{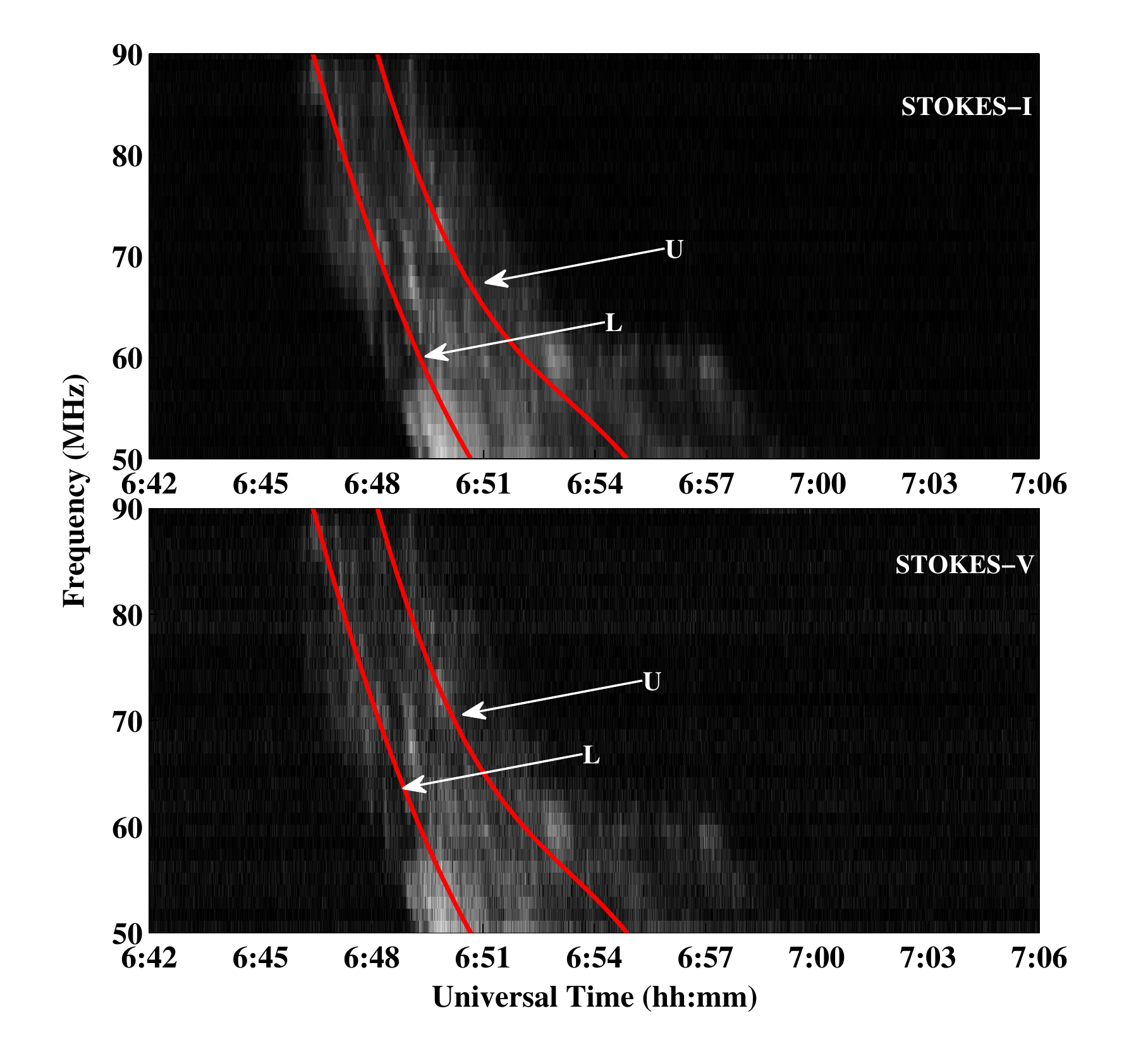}}
\caption{Dynamic spectra of the split-band type {\sc II} radio burst observed with the GRASP on 2016 March 16 during the time interval $\approx$\,06:45\,-\,07:00 UT. 
The `red' lines indicate the lower (L) and upper (U) bands of the burst.}
\label{fig:figure1}
\end{figure}

\begin{figure}[t!]
\centering
\centerline{\includegraphics[scale=0.5]{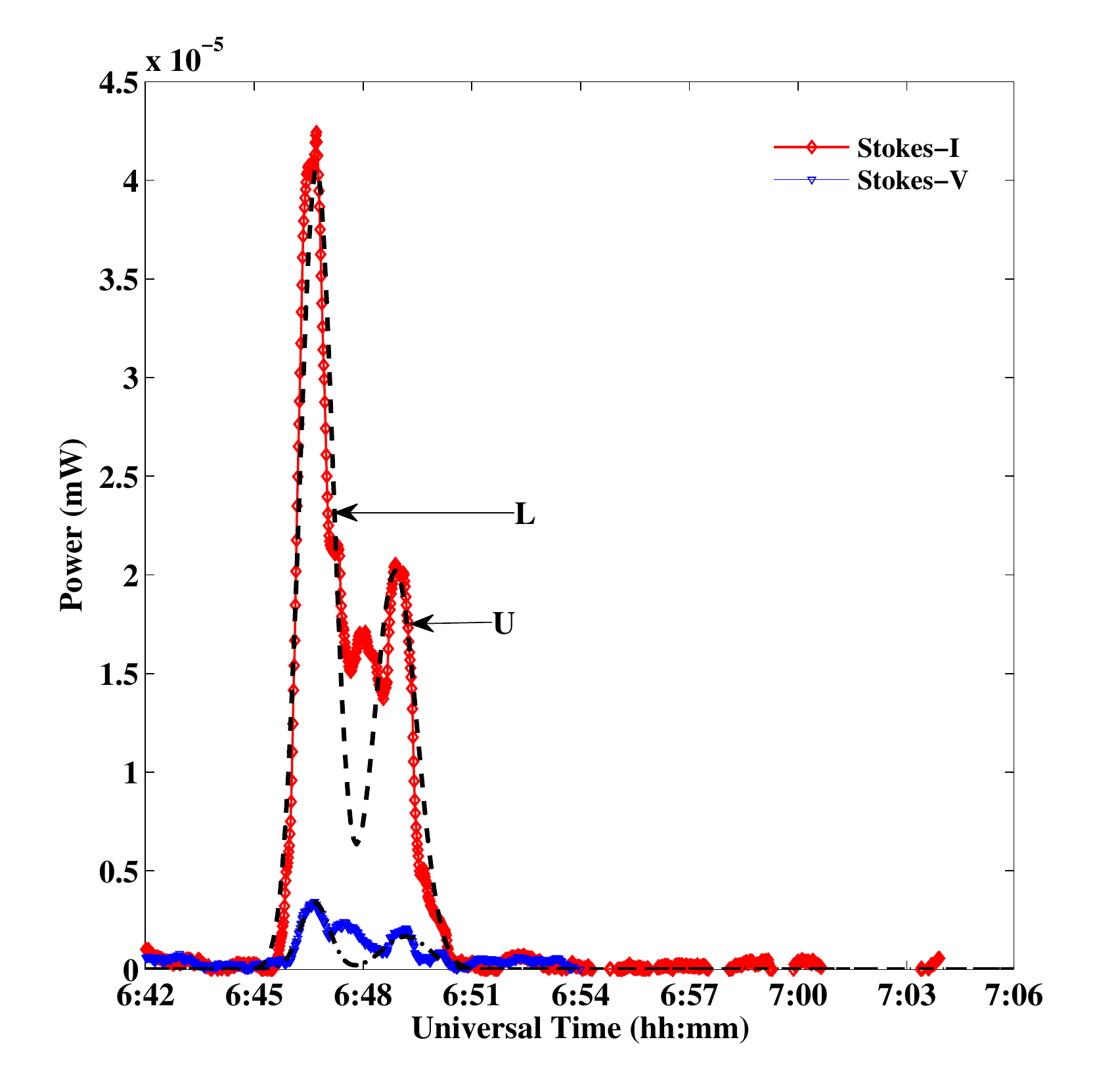}}
\caption{Temporal profile of the split-band type II burst in Figure \ref{fig:figure1} around 88 MHz, averaged over a bandwidth of $\approx$ 4 MHz. The `dotted' lines represent Gaussian fits to the observed profiles.}
\label{fig:figure2}
\end{figure}

The radio observations reported in the present work were carried out using the different facilities operated by the Indian Institute of Astrophysics (IIA) in the Gauribidanur 
observatory\footnote{\url{https://www.iiap.res.in/?q=centers/radio}} \citep{Ramesh2011a}. The \textit{Gauribidanur RAdio Spectro-Polarimeter} (GRASP; \cite{Kishore2015,Hariharan2015}) observed a split-band type II radio burst
from the Sun on 2016 March 16 during the period $\approx$\,06:45\,-\,07:00 UT.
The frequency range of the burst was $\approx$\,90\,-\,50 MHz. 
Figure \ref{fig:figure1} shows the dynamic spectra of the burst observed with the GRASP in Stokes I and V. 
Radio frequency interference in the observations are minimal \citep{Monstein2007}.
The estimated peak degree of circular polarization (\textit{dcp}) is in the range $\approx$8\,-\,11\%. 
The duration of the lower (L) and upper (U) bands of the split-band burst at a typical frequency like 88 MHz are $\approx$\,2.3 min and $\approx$\,2.5 min, respectively (see Figure \ref{fig:figure2}). 
The half-power width of the response pattern of GRASP is
$\approx\,90^{\arcdeg} \times 60^{\arcdeg}$ (right ascension, R.A. $\times$ declination, decl.) and is nearly
independent of frequency. The primary receiving element used
in GRASP is a Crossed Log-Periodic Dipole \citep{Sasi2013a}.
The integration time is $\approx$\,250\,msec, and the observing bandwidth is 
$\approx\,$1 MHz at each frequency. The antenna and the receiver systems were calibrated by carrying
out observations in the direction of the Galactic center as described in \cite{Kishore2015}.
The burst was observed elsewhere also\footnote{\url{ftp://ftp.swpc.noaa.gov/pub/warehouse/2016/}} including the \textit{Gauribidanur Radio Interferometer Polarimeter} (GRIP; \cite{Ramesh2008}), the Gauribidanur LOw-frequency Solar Spectrograph (GLOSS; \cite{Ebenezer2001,Ebenezer2007,Kishore2014})  and e-Callisto \citep{Benz2009} in Gauribidanur\footnote{\url{http://soleil.i4ds.ch/solarradio/qkl/2016/03/16/GAURI_20160316_064459_59.fit.gz.png}} and 
Ooty\footnote{\url{http://soleil.i4ds.ch/solarradio/qkl/2016/03/16/OOTY_20160316_064443_59.fit.gz.png}}. It was associated with a C2.2 class soft X-ray (SXR) flare observed with the Geostationary Operational Environmental Satellite (GOES-15) from the NOAA sunspot active region AR12522 located at the 
heliographic coordinates N12W83\footnote{\url{https://www.solarmonitor.org/index.php?date=20160316}}.
The above flare was present in the time interval $\approx$\,06:34\,-\,06:57 UT, with peak at $\approx$\,06:46 UT.
The location of the split-band burst in the solar atmosphere was inferred from observations with the \textit{Gauribidanur RAdioheliograPH} (GRAPH; \cite{Ramesh1998,Ramesh1999a,Ramesh2006b} at 80 MHz and 53.3 MHz 
(see Figure \ref{fig:figure5}). The GRAPH is a T-shaped radio interferometer array which produces two-dimensional images of the solar corona with an angular resolution of $\approx5{\arcmin} \times 7{\arcmin}$ (R.A.\,$\times$\,decl.) at a typical frequency like 80 MHz. The integration time is $\approx\,$250 msec and the observing bandwidth is $\approx\,$2 MHz. We would like to add here that both the type II bursts shown in Figure \ref{fig:figure5} correspond to the lower (L) band of the split-band type II burst in Figure \ref{fig:figure1}.

\subsection{Optical Observations}

\begin{figure}[t!]
\centering
\centerline{\includegraphics[angle=0,height=7cm,width=16cm]{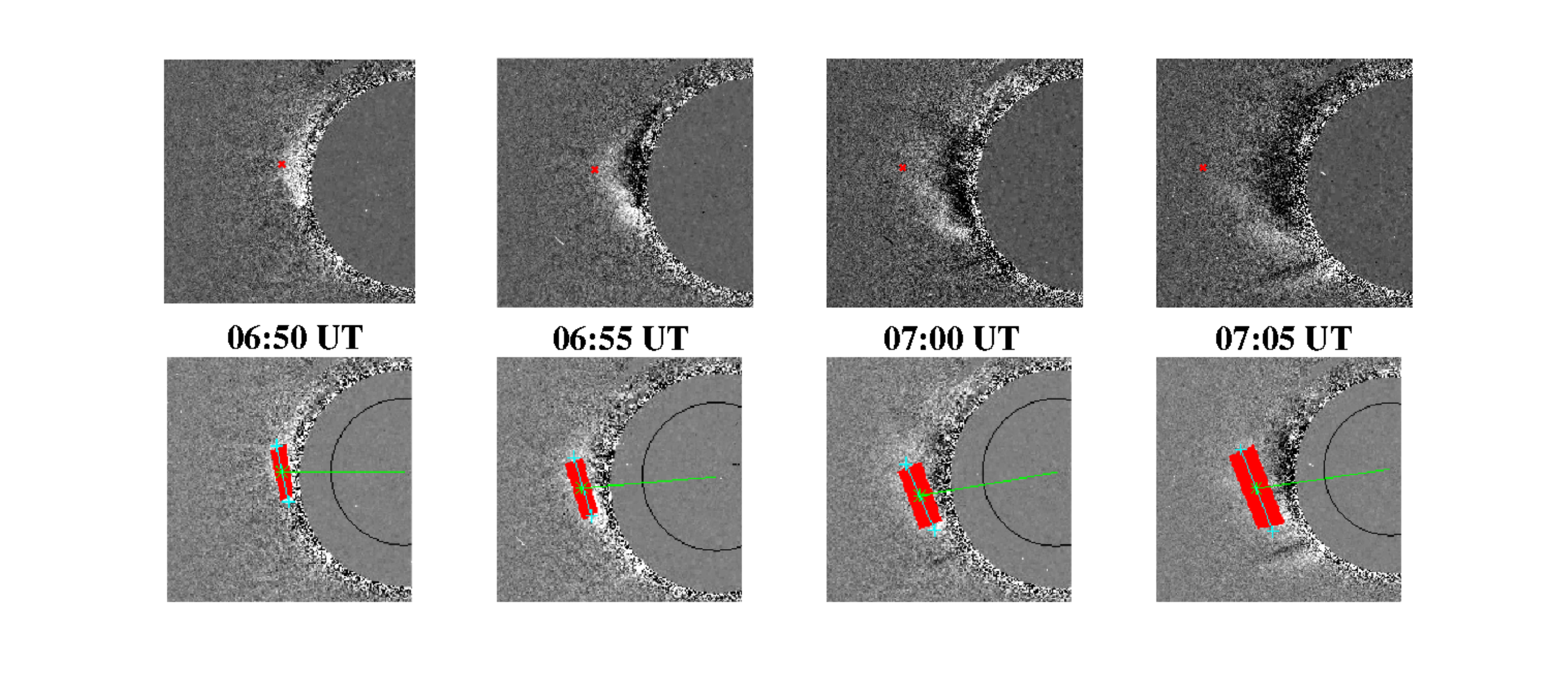}}
\caption{Upper panel: STEREO-A/COR1 pB difference images of the CME that was observed on 2016 March 16 between $\approx$\,06:50 and $\approx$ 07:05 UT. The `red' cross marks indicate the LE of the CME at different epochs. The `gray' circle represents the coronagraph occulter (radius $\approx$\,1.4$\rm R_{\odot}$). Lower panel: Same as above but with marking of the CME region (indicated by the red box) used for estimating the densities in Table \ref{tab:table1}. The `green' asterisk indicates the centroid of the CME, and the `green' line indicates its heliocentric distance. The `black' circle indicates the solar limb (radius = 1$\rm R_{\odot}$).}
\label{fig:figure3}
\end{figure}

The optical data reported in the present work were obtained  in EUV at 211{\AA} with the Atmospheric Imaging Assembly (AIA; \cite{Lemen2012}) on board the Solar Dynamics Observatory (SDO), and in whitelight with the COR1 coronoagraph of the \textit{Sun-Earth Connection Coronal and Heliospheric Investigation} (SECCHI; \cite{Howard2008}) on board the \textit{Solar Terrestrial Relationship Observatory} (STEREO)\footnote{\url{https://cor1.gsfc.nasa.gov/}} and the Large Angle and Spectrometric Coronagraph (LASCO; \cite{Brueckner1995}) on board the Solar and Heliospheric Observatory (SOHO).
The STEREO-A/COR1 instrument observed a CME around the same time as the type II burst in Figure \ref{fig:figure1}. The CME was first seen in the STEREO-A/COR1 
field-of-view (FOV) at $\approx$\,06:50 UT and was noticeable till $\approx$\,07:05 UT (see Figure \ref{fig:figure3}).
The projected heliocentric distance of the centroid of the CME ($r_{CME}$) during
its first appearance was $\rm 1.66R_{\odot}$.
The angular width of the CME is $\approx$\,$36^{\circ}$. The source region for this CME is the active region AR12522 (N12W83) mentioned in Section 2.1. 
STEREO-A was at $\approx$\,E$163^{\circ}$ during the onset of the 
CME\footnote{\url{https://stereo-ssc.nascom.nasa.gov/cgi-bin/make\_where\_gif}}. The location of the active region therefore corresponds to 
$\approx$\,$24^{\arcdeg}$ behind the limb for the STEREO-A view.

\begin{figure}[t!]
\centering
\centerline{\includegraphics[angle=0,height=5cm,width=15cm]{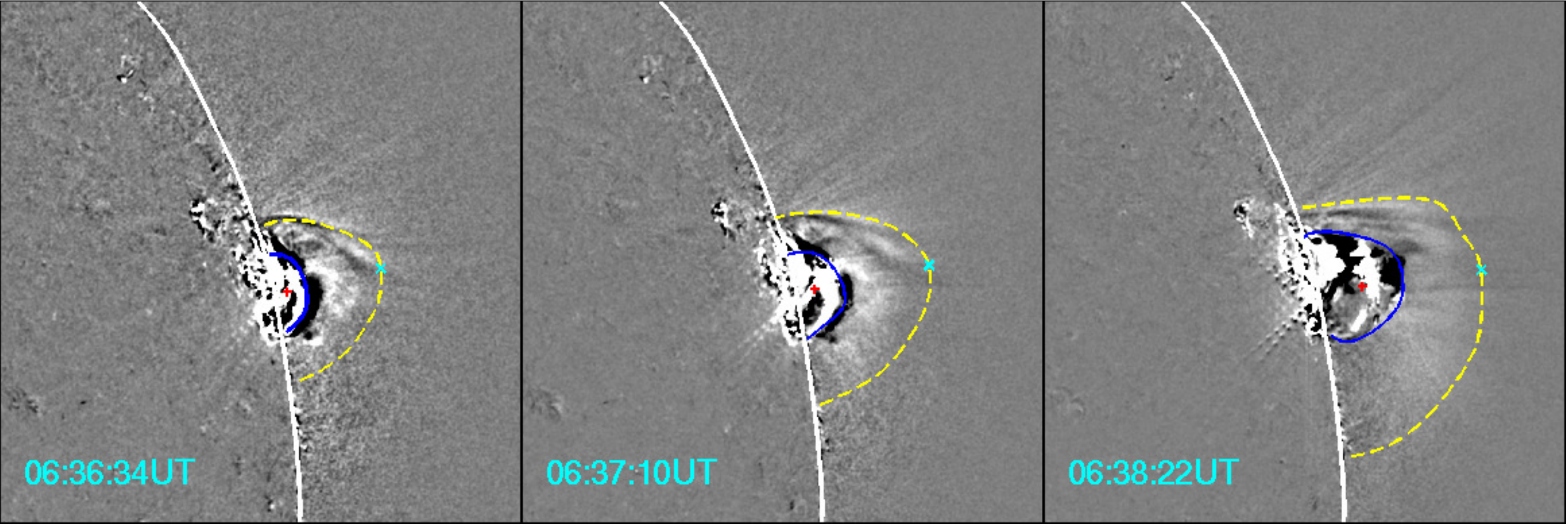}}
\caption{Evolution of the flux rope and shock in SDO/AIA 211{\AA} FOV near the source region of the CME in Figure \ref{fig:figure3}. The 'white' line indicates the solar limb (radius = 1$\rm R_{\odot}$). The `blue' and `yellow' markings indicate the flux rope structure and shock ahead of it, respectively. The `red' plus marks correspond to the centre of the hemispherical structure (assumed) for the flux rope. The `cyan' crosses represent the LE of the shock.}
\label{fig:figure4}
\end{figure}

\begin{figure}[t!]
\centering
\centerline{\includegraphics[angle=0,height=8cm,width=10cm]{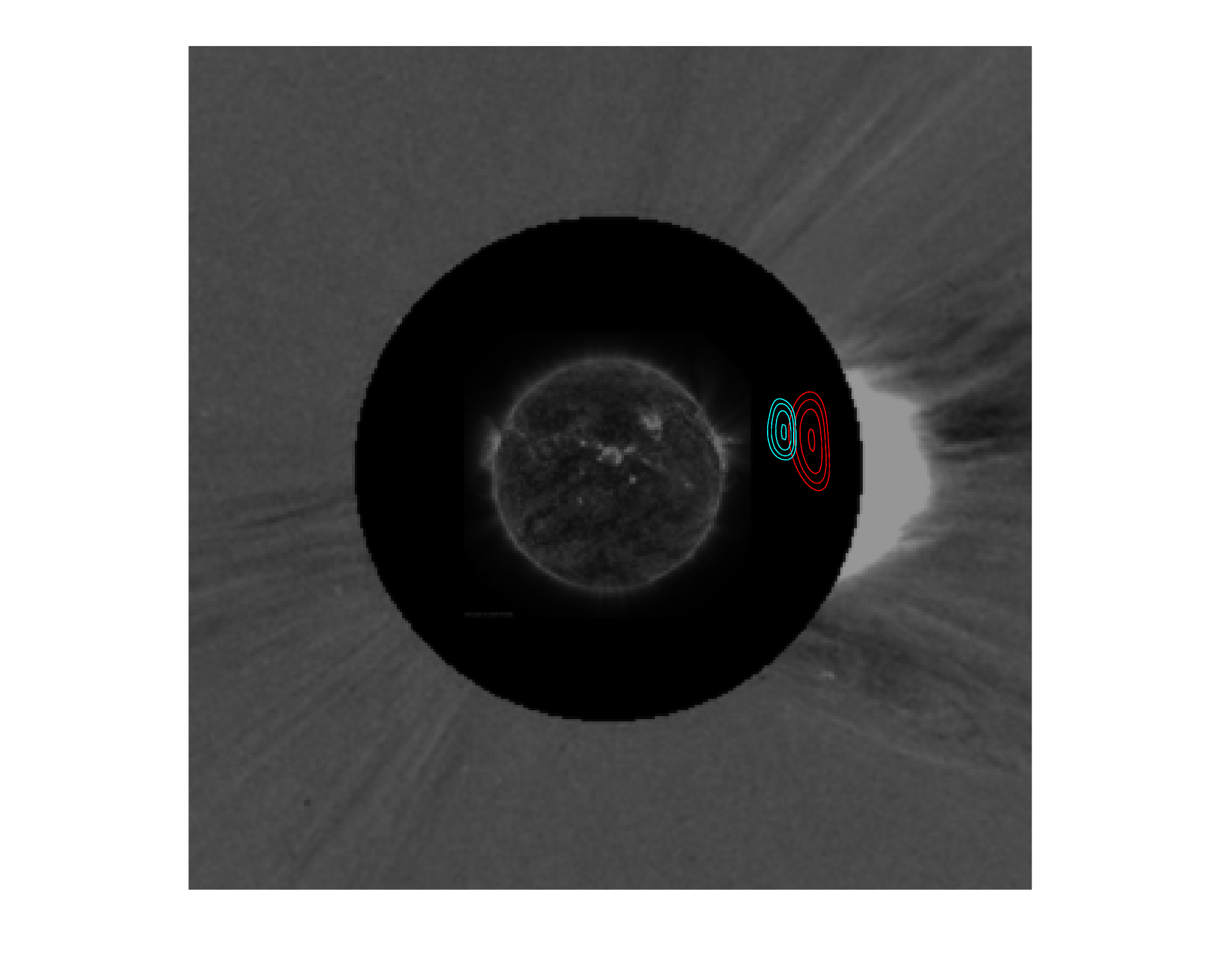}}
\caption{Locations of the type II bursts observed with the GRAPH on 2016 March 16 at 80 MHz ($\approx$\,06:47:15 UT) and 53.3 MHz ($\approx$\,06:49:48 UT) superposed on the SDO/AIA 211{\AA} image ($\approx$\,06:39:36 UT), and SOHO/LASCO-C2 difference image ($\approx$\,07:00 UT) 
obtained on the same day. Solar north is straight up and east is to 
the left. 
 The `red' and `cyan' color contours represent the GRAPH observations at 53.3 MHz and  80 MHz, respectively.
The peak brightness temperatures ($\rm T_{b}$) of the burst are $\approx$\,$2.66 \times 10^8$ K (80 MHz) and $\approx$\,$4.46 \times 10^8$ K (53.3 MHz). The radio contours shown are at 50$\%$, 65$\%$,80$\%$ and 99$\%$ of the peak $\rm T_{b}$. The `black' circle indicates the occulting disk of the coronagraph. Its radius is $\approx$\,2.2$\rm R_{\odot}$. The bright patch of emission above the coronagraph occulter on its west corresponds to the CME mentioned in the text.} 
\label{fig:figure5}
\end{figure}

The de-projected heliocentric distances of the CME were calculated for STEREO-A/COR1 images by assuming that the projection effects vary as 1/cos($\phi$), where $\phi$ is the angle from the plane-of-sky (POS) and is equal to $\approx24^{\arcdeg}$ in the present case (see Table \ref{tab:table1} for the de-projected $r_{CME}$ at different epochs). Figure \ref{fig:figure4} shows SDO/AIA 211{\AA} observations of activity in the source region of the above CME. 
The evolution of a flux rope (marked with blue line) and a diffuse shock ahead of it (marked with yellow line), as described in \citet{Gopalswamy2012}, can be clearly noticed. The leading edge (LE) of the flux rope ($r_{fl}$) and the shock
($r_{sh}$) are located at $\rm\approx1.06R_{\odot}$ and 
$\rm\approx1.13R_{\odot}$, respectively, at $\approx$06:36:36 UT.  
The values of $r_{fl}$, $r_{sh}$, and the radius of curvature ($r_{c}$) of the flux rope at different epochs are listed in Table \ref{tab:table2}.
Figure \ref{fig:figure5} shows the SOHO/LASCO-C2 observations of the CME at 
$\approx$\,07:00 UT,
along with the SDO/AIA 211{\AA} and GRAPH observations at epochs earlier to the appearance of the CME in the SOHO/LASCO-C2 FOV. It appears that the flux rope structure in EUV, the type II radio burst, and the whitelight CME are all closely associated. Note that the projection effects are very minimal in all the above three observations since AR12522 is almost at the limb of the Sun. We find that the shock is not noticeable in the STEREO-A/COR1 whitelight observations (see Figure \ref{fig:figure3}). It is possible that the shock had become fainter by the time the CME reached the STEREO-A/COR1 FOV.

\section{Analysis and Results}\label{data}
\subsection{Estimates of coronal electron density ($N_{e}$)}
\label{graph}

\subsubsection{Radio imaging observations with GRAPH}

An inspection of Figure \ref{fig:figure5} indicates that the centroid of the type {\sc II} burst ($r_{radio}$) observed with the GRAPH at 80 MHz and 53.3 MHz are located at 
$\approx$\,1.6$\pm$0.2$\rm R_{\odot}$ and $\approx$\,1.9$\pm$0.2$\rm R_{\odot}$, respectively. 
Any possible error in the position of the burst due to propagation effects such as scattering by density inhomogeneities in the solar corona and/or refraction in the Earth's ionosphere is expected to be within the above error limit \citep{Stewart1982,Ramesh1999b,Ramesh2006a,Ramesh2012b,Kathiravan2011,
Mercier2015,Mugundhan2016,Mugundhan2018}. 
The fact that the Sun is presently in the phase of minimum activity, (during which the observations reported in the present work were carried out) also indicates that the scattering will be less \citep{Sasikumar2016,Mugundhan2017}.
We calculated $N_{e}$ at the above two heliocentric distances 
using the relation $N_{e}$\,=\,$\left(\frac{f_p}{9\times10^{-3}}\right)^2$, where $f_{p}$ is the fundamental plasma frequency in units of MHz, and $N_{e}$ is in units of $\rm cm^{-3}$.
We would like to note here that the type II burst in the present case is mostly due to harmonic plasma emission ($2f_{p}$) since the locations of the bursts as observed with the GRAPH at 80 MHz and 53.3 MHz are above the limb (see Figure \ref{fig:figure5}). The consistency of the estimated peak \textit{dcp} of the bursts from the GRASP observations ($\approx$\,8\,-\,11\%, see Section 2.1) with those reported in the literature for harmonic plasma emission also indicate the same (see for example \cite{Dulk1980}). 
An inspection of the dynamic spectra of the type II burst as observed with the GLOSS indicates the presence of a faint fundamental component of the type II burst at frequencies $\lesssim$ 50 MHz\footnote{\url{https://www.iiap.res.in/gauribidanur/GLOSS-dailyimages/Mar-2016/GBD_DSPEC_20160316.jpeg}}.
These confirm that the type II bursts observed with GRASP (Figure \ref{fig:figure1}) 
and GRAPH (Figure \ref{fig:figure5}) are due to harmonic emission. 
So we substituted 40 MHz and 26.7 MHz for $f_{p}$ in the above relation, 
and obtained the values of $N_{e}$ as $1.98 \times 10^{7}~\rm cm^{-3}$ 
at $\approx$\,1.6$\rm R_{\odot}$ 
($f_{p}$ = 40 MHz) and $8.77 \times 10^{6}~\rm cm^{-3}$ at $\approx$\,1.9$\rm R_{\odot}$ ($f_{p}$ = 26.7 MHz).

\subsubsection{Whitelight observations with STEREO-A/COR1}
\label{sec3.1.2}

The pB measurements with the STEREO-A/COR1 were used to estimate the densities before the occurrence of the CME (i.e. the `background' corona at the location of the CME) and during the CME, at different heliocentric distances. The difference images used for this purpose were obtained using the observations of the CME at $\approx$\,06:50 UT, \,06:55 UT, \,07:00 UT and \,07:05 UT, and that of the `undisturbed' background corona at $\approx$\,06:45 UT (see Figure \ref{fig:figure3}). Table \ref{tab:table1}
provides the CME related details obtained from the aforesaid difference images. The de-projected $r_{CME}$ at the above epochs are listed in column 2 of Table \ref{tab:table1}. Note that we had multiplied the measured projected values of $r_{CME}$ by 1/cos(24\rm$^{\arcdeg}$) to remove the projection effects (see Section 2.2). The $N_{e}$ values of the `undisturbed' background corona and the CME at the corresponding heliocentric distances are listed in columns 3 and 4 of Table \ref{tab:table1}. The densities were calculated using the spherically symmetrical inversion technique \citep{Wang2014}. 
Note the aforesaid densities correspond to the average density inside the region enclosed by the `red' box in lower panel of Figure \ref{fig:figure3}. 

\begin{table}[!t]
\centering
\caption{Density estimates using STEREO-A/COR1 data}
\label{tab:table1}
\begin{tabular}{cccc}     
\noalign{\smallskip}\hline\noalign{\smallskip}
Time & de-projected & Background & CME \\
(UT) & $r_{CME}$\footnote{centroid of the CME} & density 
& density \\
& ($\rm R_{\odot}$) & ($\rm \times 10^{6}$ $\rm cm^{-3}$) & 
($\times 10^{6}$ $\rm cm^{-3}$)\\
\hline
06:50 & 1.82 & 7.34$\pm$1.53 & 2.71$\pm$2.46 \\
06:55 & 2.00 & 4.32$\pm$0.86 & 2.35$\pm$1.56 \\
07:00 & 2.06 & 3.49$\pm$0.90 & 1.66$\pm$1.17 \\
07:05 & 2.24 & 2.21$\pm$0.73 & 0.95$\pm$0.91 \\
{\bf 1} & {\bf 2} & {\bf 3} & {\bf 4} \\
\hline
\end{tabular}
\end{table}

\begin{figure}[t!]
\centering
\centerline{\includegraphics[angle=0,height=8cm,width=8cm]{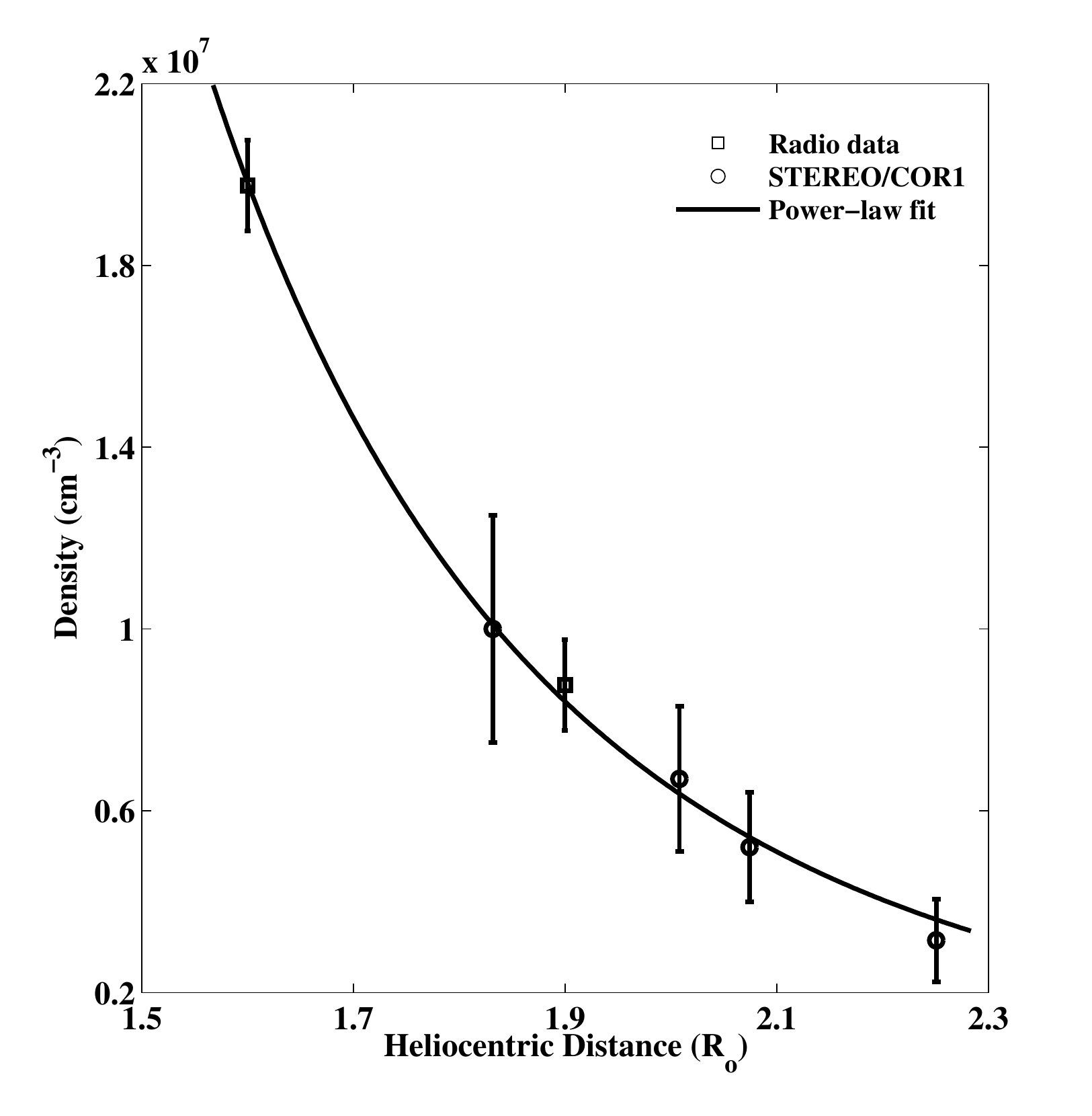}}
\caption{Density estimates from radio (GRAPH) and white-light (STEREO-A/COR1) observations. The solid line is a power-law fit ($N_{e}(r)$\,=\,$2.3\times10^8\,r^{-5.3}$) to the data.}
\label{fig:figure6}
\end{figure}

Figure \ref{fig:figure6} shows the plot of the $N_{e}$ values obtained using GRAPH and STEREO-A/COR1 observations as mentioned above. The error in the density estimates from STEREO-A/COR1 is chiefly due to the errors associated with the instrumental background subtraction, and the spherically symmetric approximation \citep{Wang2014,Wang2017}. The error in the density estimates from GRAPH is due to variation in $N_e$ over the bandwidth of observations ($\approx$2 MHz).
The power-law fit to the data  indicate that $N_{e}(r)$\,=\,$2.3\times10^8\,r^{-5.3}$ in the range $r\approx$1.6\,-\,2.2$\rm R_{\odot}$. 
Note that $N_{e}(r)$ varies typically as $r^{-6}$ in the range 
$1.1 \lesssim r \lesssim 2.3\rm R_{\odot}$ \citep{Baumbach1937}. Considering this, and  
since we are interested in understanding the characteristics of the CME close to the Sun also in the present case 
using the SDO/AIA 211\AA\ observations of the associated flux rope structure (see Figure \ref{fig:figure4}), we assumed that the above empirical relationship should be valid over
$r\approx$1.1\,-\,2.2$\rm R_{\odot}$.
We find that $N_{e}(r)$ estimated using the above relation for the SDO/AIA 211\AA\ observations in Figure \ref{fig:figure4} are reasonably consistent with the $N_{e}(r)$ values reported by \citet{Zucca2014_1} in the same distance range ($r\approx$1.1\,-\,1.3$\rm R_{\odot}$) utilizing  the emission measures derived from SDO/AIA observations for a similar flare associated CME/type II burst event.

\subsection{Tracing the path of the CME}
\label{HT}

\begin{figure}[t!]
\centering
\centerline{\includegraphics[angle=0,height=7cm,keepaspectratio]{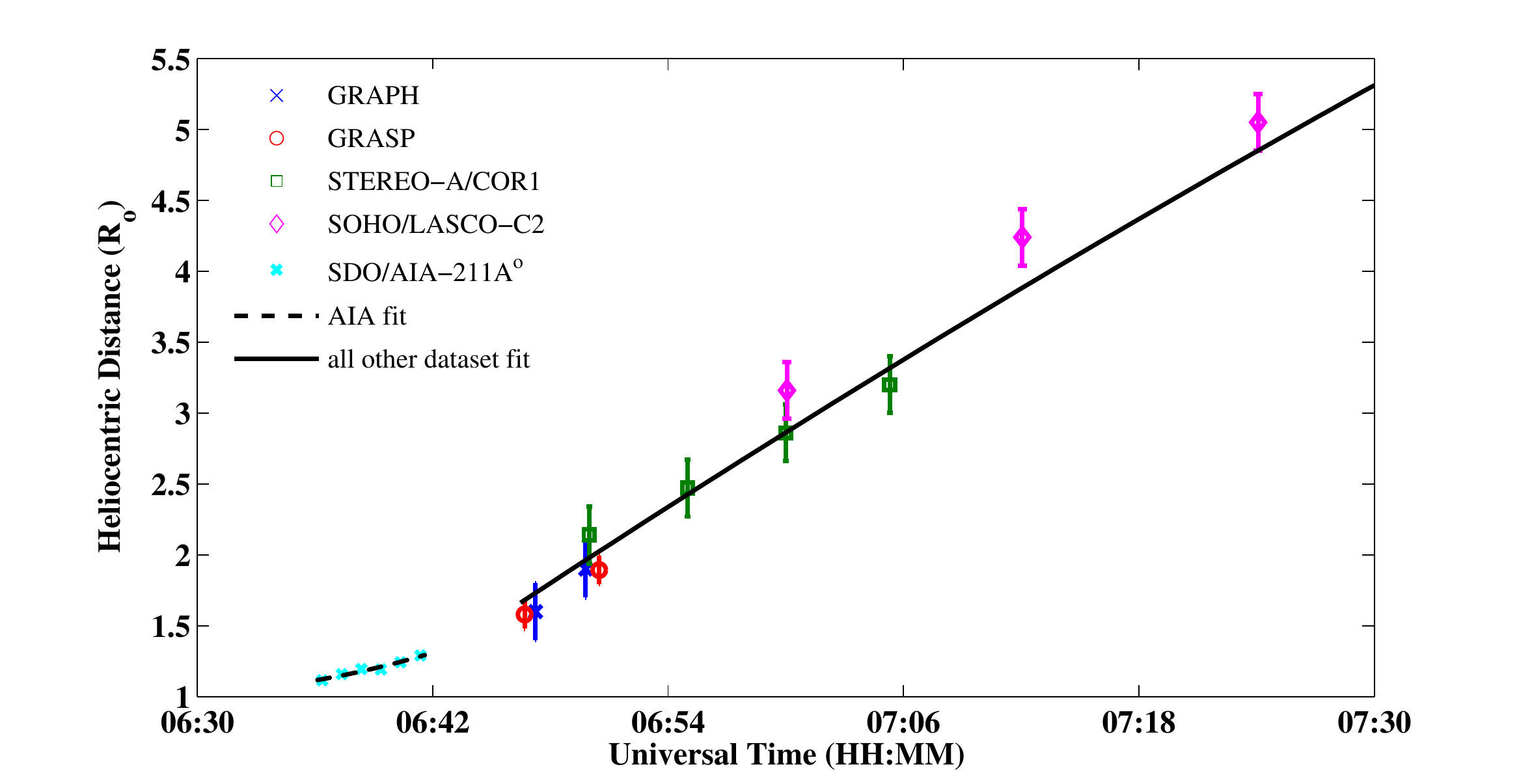}}
\caption{Height-time plot
of the EUV shock (SDO/AIA 211\AA), type II radio bursts
(GRAPH and GRASP), and the whitelight CME (STEREO-A/COR1 and SOHO/LASCO-C2). 
The `dashed' black line is a quadratic fit to SDO/AIA {211\AA} data, and the `solid' black line is a quadratic fit to the GRAPH, GRASP, STEREO-A/COR1 and SOHO/LASCO-C2 data.}
\label{fig:figure7}
\end{figure}

Figure \ref{fig:figure7} shows the height-time (h-t) plot of 
the leading edge (LE) of the EUV flux rope structure as observed with SDO/AIA 
211{\AA}, locations of the type II bursts observed with the GRAPH at 80 and 53.3 MHz and GRASP at two different frequencies in the range $\approx$\,90\,-\,50 MHz, the LE of the CME in the FOV of the STEREO-A/COR1 coronagraph (see Figure \ref{fig:figure3}) and SOHO/LASCO-C2. 
For the GRASP data in the plot we used two representative frequencies, viz. 82 and 50 MHz (in the lower band $L$ of the harmonic emission, see Figure \ref{fig:figure1}). 
Their heliocentric distances are
$r \approx$\, 1.58$\rm R_{\odot}$ and 1.93$\rm R_{\odot}$, respectively. 
Note that in the case of the GRASP observations, the locations of 
the type II bursts were estimated using the relationship between $f_{p}$ and $N_{e}$, and the model for $N_{e}(r)$ derived in Section 3.1. The SDO/AIA  211{\AA} values are limited to less than $r \approx$\, 1.23$\rm R_{\odot}$ and radio + whitelight values are available only 
beyond $r \approx$\, 1.58$\rm R_{\odot}$.
So we used two separate quadratic fits for the h-t data in Figure \ref{fig:figure7}: one for the former with an acceleration of $\rm \approx 1259~m/s^{2}$, 
and the other for the latter with an acceleration of $\rm \approx - 46 ~m/s^{2}$.
The comparatively large acceleration in the SDO/AIA 211{\AA} FOV was during the onset-peak phase of the associated GOES/SXR flare (see Section 2.1).
This is consistent with earlier reports of acceleration of the flux rope structure in the SDO/AIA observations during the impulsive phase of the flare (see for example \citet{Zhang2012}). The decrease in acceleration in the present case is during the decay phase of the flare.
We find that there is reasonable consistency between the two quadratic fits
in Figure \ref{fig:figure7}.
This is expected since the early signature of a CME close to the Sun 
is usually an expanding flux rope structure 
\citep{Pomoell2009,Ma2011,Patsourakos2012,Gopalswamy2012,Gopalswamy2013,
Cho2013}, and CME driven magnetohydrodynamic (MHD) shocks generate type II bursts in the solar atmosphere \citep{Mann1995,Aurass1997,Clasen2002,Gopalswamy2006,Cho2008,Vrsnak2008,Gopalswamy2009,
Ramesh2010b,Ramesh2012a,Anshu2017a}.
An estimate of the linear speed of the CME LE from the whitelight data 
(STEREO-A/COR1 and SOHO/LASCO-C2) in the range $r \approx$\,2.0-5.0\,$\rm R_{\odot}$  
(see Figure \ref{fig:figure7}) indicates
it is $\approx$ 1000 km/s. The estimated shock speed ($v_{s}=v_a \times M_a$, where
$v_a$ is the Alf\'{v}en speed and $M_a$ is the Mach number)
for the type II burst is $\approx$ 825 km/s (see Table \ref{tab:table3}).
This is in good agreement with the speed of the CME LE.
Note that though a shock was observed in the SDO/AIA 211{\AA} FOV, 
no type II burst was observed at that time. One likely reason for the absence of the type II burst could be the smaller values of $M_a$
associated with the above shock (see Table \ref{tab:table2}). According to  \citet{mann2003formation,warmuth2005model}, $M_a$ should well exceed 
unity ($\gtrsim 1.4$) for the occurrence of type II burst.

\subsection{Estimates of the coronal magnetic field strength ($B$)}
\label{B}

Our aim is to directly estimate $B$ using the observed data and with minimal assumptions. 
We used the following theoretical relation for this purpose:
\begin{equation}
B=\frac{v_a \times \sqrt{N_{e}}}{2.18\times10^6}
\end{equation}
where $B$ is in units of G.
We used the empirical relationship in Section 3.1 to obtain $N_{e}(r)$. The estimated values are in the range $\approx$\,$1.39 \times 10^8$\,-\,$3.6 \times 10^6 \rm cm^{-3}$ over $r \approx$\,1.10\,-\,2.20$\rm R_{\odot}$, the combined distance range of the SDO/AIA 211{\AA} and radio observations in the present case. $M_{{a}}$ was estimated independently for the aforementioned two observations since they correspond to different heliocentric distance ranges.

\subsubsection{SDO/AIA 211\AA\ observations}

\begin{table}
\centering
\caption{Estimates of $B$ and the related parameters from SDO/AIA 211{\AA} observations}
\label{tab:table2}
\begin{tabular}{ccccccccc }     
\noalign{\smallskip}\hline\noalign{\smallskip}
Time & $r_{sh}$ & $r_{fl}$ & $r_{c}$ &  $\Delta r$ & $\delta$ & $\rm M_{a}$  & $v_a$ & B \\
(UT) & ($\rm R_{\odot}$) & ($\rm R_{\odot}$) & ($\rm R_{\odot}$) & ($\rm R_{\odot}$) & & & (km/s) & (G) \\
\hline

06:36:34 & 1.12 & 1.04 & 0.025 & 0.083 & 3.35 & 1.12 & --  & --\\
06:37:10 & 1.15 & 1.06 & 0.035 & 0.090 & 2.59 & 1.15 & 401 & 1.93\\
06:37:46 & 1.17 & 1.08 & 0.040 & 0.101 & 2.50 & 1.16 & 400 & 1.83\\
06:38:22 & 1.19 & 1.10 & 0.046 & 0.095 & 2.06 & 1.19 & 390 & 1.74\\

{\bf 1} & {\bf 2} & {\bf 3} & {\bf 4} & {\bf 5} & {\bf 6} & {\bf 7} & {\bf 8} & {\bf 9}  \\
\hline
\end{tabular}
\end{table}

Figure \ref{fig:figure4} shows the initial stages of the CME formation in the SDO/AIA 211{\AA} FOV in the present case. 
Measuring the locations and characteristics of the corresponding structures, i.e. the flux rope and the shock ahead of it, at different epochs help to calculate $M_{{a}}$ using the relation (see for example \citet{Gopalswamy2012, veronig2010}),  
\begin{equation}
M_a=\sqrt{1+[1.24\delta-(\gamma-1)/(\gamma+1)]^{-1}}
\end{equation}
where $\delta$ is the relative stand-off distance and $\gamma$ is the adiabatic constant. The heliocentric distance of the shock ($r_{sh}$), LE of the CME flux rope ($r_{fl}$), thickness of the shock
$\Delta r$\,=\,$r_{sh}$\,-\,$r_{fl}$, and radius of curvature 
($r_{c}$) of the CME flux rope are used to calculate $\delta$ (=$\frac{\Delta r}{r_{c}}$). 
$\gamma$ was assumed to be 4/3 for the present calculations (see \cite{Anshu2017b,Anshu2017c} for details). The different values estimated using Figure \ref{fig:figure4} are listed in columns 2-6 of Table \ref{tab:table2}. 
We then calculated $v_{s}$ for the adjacent time intervals in column 1 using the values of $r_{sh}$ in column 2. Finally, $v_a$ values in column 8 were obtained using the relation $v_{a}=v_{s}/M_{a}$. 
We find that the location of the active region in the 
present work and that of the event reported in \citet{Gopalswamy2012} are nearly the same ($\approx$ W84). Furthermore, the $v_{a}$ values ($\approx 400-500$ km/s)
and the angular width of the CME ($\approx 36{\arcdeg}$) are also 
reasonably close in the two cases. So, assuming 06:36:34 UT as the first appearance time (i.e. $t$ = 0 of the flux rope and the shock in Figure \ref{fig:figure4}), we independently calculated the corresponding  
$r_{sh}$, $r_{fl}$, and $r_{c}$ values as a function of time 
using the empirical equations mentioned in Figures 3a and 3b of \citet{Gopalswamy2012}.
The constants in the aforementioned equations were replaced by the values of  
$r_{sh}$, $r_{fl}$, and $r_{c}$ at 06:36:34 UT (see Table \ref{tab:table2}).
Interestingly, the empirically calculated values agree well with the direct estimates. 

\subsubsection{Radio spectral observations with GRASP}
For the radio observations, $M_{{a}}$ was calculated using the following 
equation \citep{Smerd1974,Mann1995,Vrsnak2002}, 
\begin{equation}
M_a=\sqrt{\frac{X(X+5)}{2(4-X)}}
\end{equation}
where X is density jump across the shock during the type {\sc II} burst. The density jump is calculated from the instantaneous bandwidth (BDW) of the burst, i.e. $\rm BDW$\,=\,$\rm \frac{F_{U}-F_{L}}{F_{L}}$ and $\rm X$\,=\,$\rm (BDW+1)^2$. $\rm F_{U}$ and $\rm F_{L}$ are the upper and lower frequency components of the type {\sc II} burst in the dynamic spectra. 
To estimate the $B$ values, $\rm F_{L}$ is used as it corresponds to the `undisturbed' corona. 
Table \ref{tab:table3} lists the different values estimated from the type {\sc II} burst observations in Figure \ref{fig:figure1}. The $v_a$ values in column 8 were obtained 
in the same manner as the SDO/AIA 211\AA\ case described in Section 3.3.1, but equation (3) was used for the calculations of $M_{a}$.

\subsubsection{The radial variation of the coronal magnetic field strength}

\begin{figure}[t!]
\centering
\centerline{\includegraphics[angle=0,height=8cm,width=8cm]{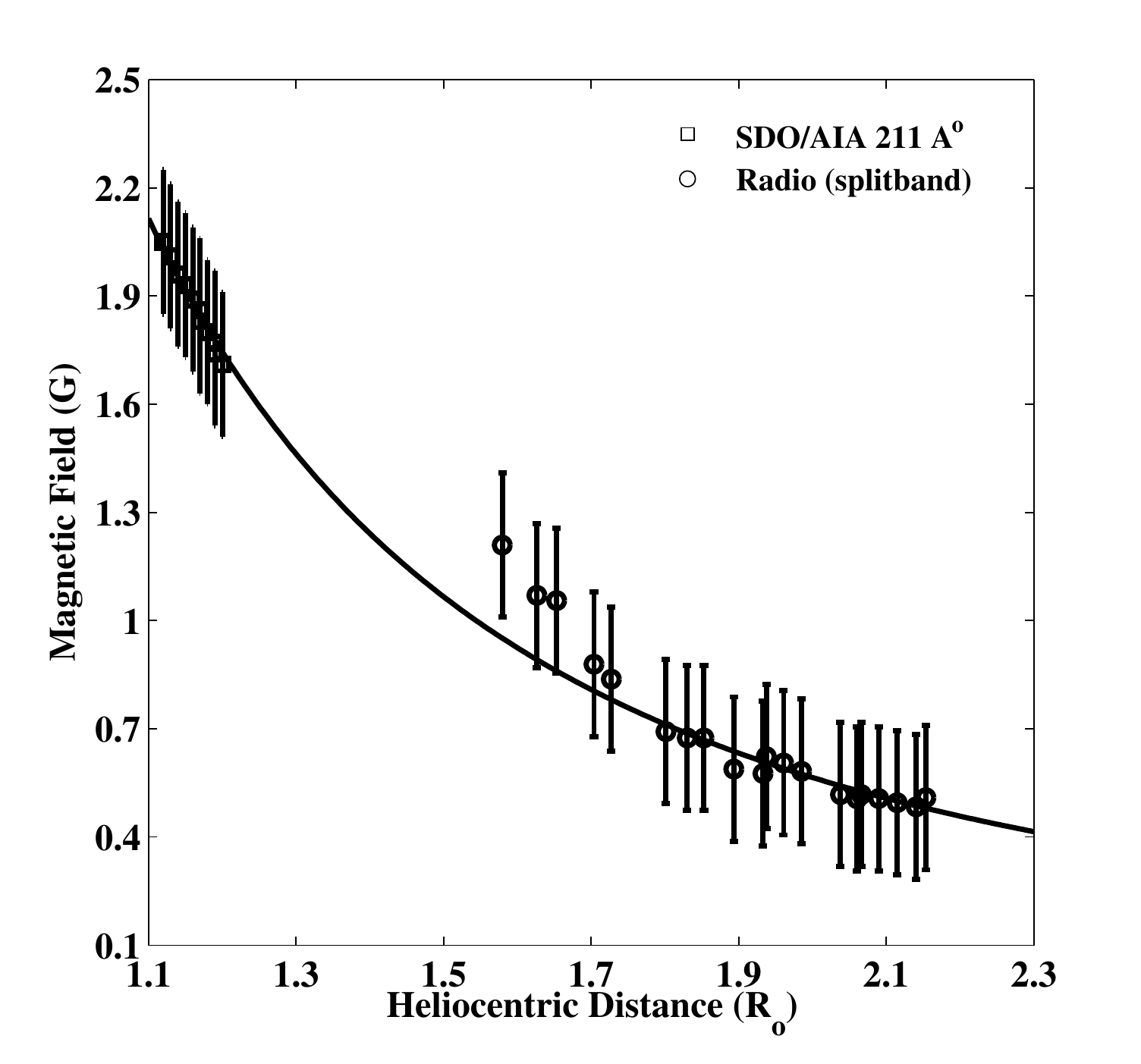}}
\caption{Estimates of $B$ from SDO/AIA 211{\AA} and radio observations. The `solid' black line is a power-law fit ($B = 2.61 \times r^{-2.21}$) to the data points.}
\label{fig:figure8}
\end{figure}

Figure \ref{fig:figure8} shows the $B$ values estimated using the SDO/AIA 211\AA\ and GRASP observations. The respective estimates are consistent with each other though they correspond to two different heliocentric distance ranges. A single power-law fit of the form $B(r)$\,=\,$2.61 \times r^{-2.21}$ nicely describes the distribution. The only available two-dimensional magnetic field map obtained using coronal Zeeman magnetometry and full-Stokes spectropolarimetric measurements indicate that $B \approx$ 3.6 G at $r \approx 1.1 \rm R_{\odot}$ \citep{Lin2004}.  
Compared to this, the present results predict $B \approx$ 2.1 G at the same distance.

\begin{table}[!t]
\centering
\caption{Estimates of $B$ and the related parameters from GRASP observations}
\label{tab:table3}
\begin{tabular}{ccccccccc}     
\noalign{\smallskip}\hline\noalign{\smallskip}
Time & $\rm F_{U}$ & $\rm F_{L}$ & BDW & X & $\rm M_{a}$ & R & $v_a$ & B \\
(UT) & (MHz)       & (MHz)       &     &  &  & ($\rm R_{\odot}$) & (km/s) & (G)\\
\hline
06:47:10 & 102.44 & 81.89 & 0.25 & 1.56 & 1.45 &  1.58 & 579 & 1.21 \\
06:48:02 & 91.12 & 72.37 & 0.26 & 1.59 & 1.47  & 1.65 &  571 & 1.06 \\
06:49:28 & 79.51 & 57.48 & 0.38 & 1.91 & 1.78  & 1.86 & 472 & 0.69 \\
06:51:00 & 65.82 & 47.65 & 0.38 & 1.91 & 1.77  & 1.99 & 473 & 0.58 \\
06:53:35 & 54.50 & 40.21 & 0.36 & 1.84 & 1.70  & 2.12 & 493 & 0.51 \\
06:57:00 & 45.57 & 35.74 & 0.27 & 1.63 & 1.51  & 2.15& 558 & 0.50 \\
{\bf 1} & {\bf 2} & {\bf 3} & {\bf 4} & {\bf 5} & {\bf 6} & {\bf 7} & {\bf 8} & {\bf 9}\\
\hline
\end{tabular}
\end{table}

\section{Summary}
\label{dis}

We have reported a CME, coronal type II radio burst and flux rope structure (in EUV) that were observed simultaneously on 2016 March 16. The radio burst was observed in both the imaging and spectral mode. The combined h-t plot indicates that all the three events are closely associated.
We derived an empirical relation for the coronal electron density ($N_{e}(r)$\,=\,$2.3\times10^8 r^{-5.3}$) using EUV observations of the flux rope structure associated with the CME, spectral and imaging observations of the type II burst associated with the CME, and pB measurements of the corresponding whitelight CME. Using the density values thus obtained along with the Alf\'{v}en Mach number ($M_{a}$) values from EUV and radio observations, we independently estimated the coronal magnetic field strength ($B(r)$). Our results indicate that $B(r)$\,=\,$2.61 \times r^{-2.21}$ in the distance range $r\approx$\,1.1\,-\,2.2$\rm R_{\odot}$. 
\citet{Mancuso2013_1} had derived $B(r)$\,=\,$3.76\times r^{-2.29}$ 
in the distance range $r\approx$\,1.8\,-\,14$\rm R_{\odot}$ by combining split-band type II 
observations and Faraday rotation measurements of extragalactic radio 
sources occulted by the solar corona. 
This is nearly same as the empirical relation for $B(r)$ in the present case. 
The present measurements are also in reasonable agreement with that reported 
by \citet{Lin2004} at $r\approx$\,1.1$\rm R_{\odot}$ using whitelight observations.
The consistency between the different measurements, though they correspond to different active regions observed at different epochs, strengthens the robustness of the estimates using radio observations.
We expect that the density model independent `direct' estimates of $B(r)$ reported in this work would lead to similar attempts in the future for unambiguous estimates of $B(r)$ in the region of the corona where whitelight observations are presently difficult.


We would like to thank the staff of the Gauribidanur observatory for their
help in maintenance of the antenna receiver  systems and  the  observations.
The work of TJW was supported by NASA Cooperative Agreement NNG11PL10A to CUA and NASA grants 80NSSC18K1131 and 80NSSC18K0668. NG was supported in part by NASA's Living With a Star program. AK acknowledges V. Mugundhan and Vaibhav Pant for discussions.
The SOHO data are produced by a consortium of the Naval Research Laboratory (USA),
Max-Planck-Institut fuer Aeronomie (Germany),~Laboratoire
d'Astronomie (France),~and the University of Birmingham (UK). SOHO
is a project of international cooperation between ESA and NASA.
The SOHO-LASCO CME catalog is generated and maintained at the CDAW Data Center
by NASA and the Catholic University of America in cooperation with the
Naval Research Laboratory. 
The SDO/AIA data are courtesy of the NASA/SDOand the AIA science teams.
We thank the referee for his/her insightful report which helped us improve the paper substantially.
\\


\begin{thebibliography}{}
\expandafter\ifx\csname natexlab\endcsname\relax\def\natexlab#1{#1}\fi

\bibitem[{{Aurass}(1997)}]{Aurass1997}
{Aurass}, H. 1997, in Lecture Notes in Physics, (Berlin:Springer-Verlag), Vol.
  483, Coronal Physics from Radio and Space Observations, ed. G.~{Trottet}, 135

\bibitem[{{Bastian} {et~al.}(2001){Bastian}, {Pick}, {Kerdraon}, {Maia}, \&
  {Vourlidas}}]{Bastian2001}
{Bastian}, T.~S., {Pick}, M., {Kerdraon}, A., {Maia}, D., \& {Vourlidas}, A.
  2001, \apjl, 558, L65

\bibitem[{{Baumbach}(1937)}]{Baumbach1937}
{Baumbach}, S. 1937, Astron. Nach, 263, 121

\bibitem[{{Benz} {et~al.}(2009){Benz}, {Monstein}, {Meyer}, {Manoharan},
  {Ramesh}, {Altyntsev}, {Lara}, {Paez}, \& {Cho}}]{Benz2009}
{Benz}, A.~O., {Monstein}, C., {Meyer}, H., {et~al.} 2009, Earth Moon and
  Planets, 104, 277

\bibitem[{{Brueckner} {et~al.}(1995){Brueckner}, {Howard}, {Koomen},
  {Korendyke}, {Michels}, {Moses}, {Socker}, {Dere}, {Lamy}, {Llebaria},
  {Bout}, {Schwenn}, {Simnett}, {Bedford}, \& {Eyles}}]{Brueckner1995}
{Brueckner}, G.~E., {Howard}, R.~A., {Koomen}, M.~J., {et~al.} 1995, \solphys,
  162, 357

\bibitem[{{Cho} {et~al.}(2008){Cho}, {Bong}, {Moon}, {Dryer}, {Shanmugaraju},
  {Lee}, \& {Park}}]{Cho2008}
{Cho}, K.-S., {Bong}, S.-C.~{Kim}, Y.-H., {Moon}, Y.-J., {et~al.} 2008, \aap,
  491, 873

\bibitem[{{Cho} {et~al.}(2013){Cho}, {Gopalswamy}, {Kwon}, {Kim}, \&
  {Yashiro}}]{Cho2013}
{Cho}, K.-S., {Gopalswamy}, N., {Kwon}, R.-Y., {Kim}, R.-S., \& {Yashiro}, S.
  2013, \apj, 765, 148

\bibitem[{{Cho} {et~al.}(2007){Cho}, {Lee}, {Gary}, {Moon}, \&
  {Park}}]{Cho2007}
{Cho}, K.-S., {Lee}, J., {Gary}, D.~E., {Moon}, Y.-J., \& {Park}, Y.-D. 2007,
  \apj, 665, 799

\bibitem[{{Cla{\ss}en} \& {Aurass}(2002)}]{Clasen2002}
{Cla{\ss}en}, H.~T., \& {Aurass}, H. 2002, \aap, 384, 1098

\bibitem[{{Dulk} \& {McLean}(1978)}]{Dulk1978}
{Dulk}, G.~A., \& {McLean}, D.~J. 1978, \solphys, 57, 279

\bibitem[{{Dulk} \& {Suzuki}(1980)}]{Dulk1980}
{Dulk}, G.~A., \& {Suzuki}, S. 1980, \aap, 88, 203

\bibitem[{{Ebenezer} {et~al.}(2001){Ebenezer}, {Ramesh}, {Subramanian},
  {Sundara Rajan}, \& {Sastry}}]{Ebenezer2001}
{Ebenezer}, E., {Ramesh}, R., {Subramanian}, K.~R., {Sundara Rajan}, M.~S., \&
  {Sastry}, C.~V. 2001, \aap, 367, 1112

\bibitem[{{Ebenezer} {et~al.}(2007){Ebenezer}, {Subramanian}, {Ramesh},
  {Sundara Rajan}, \& {Kathiravan}}]{Ebenezer2007}
{Ebenezer}, E., {Subramanian}, K.~R., {Ramesh}, R., {Sundara Rajan}, M.~S., \&
  {Kathiravan}, C. 2007, Bull. Astron. Soc. India, 35, 111

\bibitem[{{Gopalswamy}(2006)}]{Gopalswamy2006}
{Gopalswamy}, N. 2006, in Solar Eruptions and Energetic Particles, ed.
  N.~{Gopalswamy}, R.~{Mewaldt}, \& J.~{Torsti}, Geophysical Monograph Series
  Vol.165, 207

\bibitem[{{Gopalswamy} {et~al.}(2012){Gopalswamy}, {Nitta}, {Akiyama},
  {M{\"a}kel{\"a}}, \& {Yashiro}}]{Gopalswamy2012}
{Gopalswamy}, N., {Nitta}, N., {Akiyama}, S., {M{\"a}kel{\"a}}, P., \&
  {Yashiro}, S. 2012, \apj, 744, 72

\bibitem[{{Gopalswamy} {et~al.}(1986){Gopalswamy}, {Thejappa}, {Sastry}, \&
  {Tlamicha}}]{Gopalswamy1986}
{Gopalswamy}, N., {Thejappa}, G., {Sastry}, C.~V., \& {Tlamicha}, A. 1986,
  Bull. Astron. Inst. Czech., 37, 115

\bibitem[{{Gopalswamy} {et~al.}(2013){Gopalswamy}, {Xie}, {Yashiro}, \&
  {Akiyama}}]{Gopalswamy2013}
{Gopalswamy}, N., {Xie}, H.~{M\"{a}kel\"{a}}, P., {Yashiro}, S., \& {Akiyama},
  S. 2013, Adv. Sp. Res., 51, 1981

\bibitem[{{Gopalswamy} {et~al.}(2009){Gopalswamy}, {Thompson}, {Davila},
  {Kaiser}, {Yashiro}, {M\"{a}kel\"{a}}, {Michalek}, {Bougeret}, \&
  {Howard}}]{Gopalswamy2009}
{Gopalswamy}, N., {Thompson}, W.~T., {Davila}, J.~M., {et~al.} 2009, \solphys,
  259, 227

\bibitem[{{Hariharan} {et~al.}(2015){Hariharan}, {Ramesh}, \&
  {Kathiravan}}]{Hariharan2015}
{Hariharan}, K., {Ramesh}, R., \& {Kathiravan}, C. 2015, \solphys, 290, 2479

\bibitem[{{Hariharan} {et~al.}(2014){Hariharan}, {Ramesh}, {Kishore},
  {Kathiravan}, \& {Gopalswamy}}]{Hariharan2014}
{Hariharan}, K., {Ramesh}, R., {Kishore}, P., {Kathiravan}, C., \&
  {Gopalswamy}, N. 2014, \apj, 795, 14

\bibitem[{{Howard} {et~al.}(2008){Howard}, {Moses}, {Vourlidas}, {Newmark},
  {Socker}, {Plunkett}, {Korendyke}, {Cook}, {Hurley}, {Davila}, {Thompson},
  {St Cyr}, {Mentzell}, {Mehalick}, {Lemen}, {Wuelser}, {Duncan}, {Tarbell},
  {Wolfson}, {Moore}, {Harrison}, {Waltham}, {Lang}, {Davis}, {Eyles},
  {Mapson-Menard}, {Simnett}, {Halain}, {Defise}, {Mazy}, {Rochus}, {Mercier},
  {Ravet}, {Delmotte}, {Auchere}, {Delaboudiniere}, {Bothmer}, {Deutsch},
  {Wang}, {Rich}, {Cooper}, {Stephens}, {Maahs}, {Baugh}, {McMullin}, \&
  {Carter}}]{Howard2008}
{Howard}, R.~A., {Moses}, J.~D., {Vourlidas}, A., {et~al.} 2008, \ssr, 136, 67

\bibitem[{{Kathiravan} {et~al.}(2011){Kathiravan}, {Ramesh}, {Indrajit V.
  Barve}., \& {Rajalingam}}]{Kathiravan2011}
{Kathiravan}, C., {Ramesh}, R., {Indrajit V. Barve}., \& {Rajalingam}, M. 2011,
  \apj, 730, 91

\bibitem[{{Kishore} {et~al.}(2017){Kishore}, {Kathiravan}, {Ramesh}, \&
  {Ebenezer}}]{Kishore2017}
{Kishore}, P., {Kathiravan}, C., {Ramesh}, R., \& {Ebenezer}, E. 2017, J.
  Astrophys. Astron., 38, 24

\bibitem[{{Kishore} {et~al.}(2014){Kishore}, {Kathiravan}, {Ramesh},
  {Rajalingam}, \& {Indrajit V. Barve}}]{Kishore2014}
{Kishore}, P., {Kathiravan}, C., {Ramesh}, R., {Rajalingam}, M., \& {Indrajit
  V. Barve}. 2014, \solphys, 289, 3995

\bibitem[{{Kishore} {et~al.}(2016){Kishore}, {Ramesh}, {Hariharan},
  {Kathiravan}, \& {Gopalswamy}}]{Kishore2016}
{Kishore}, P., {Ramesh}, R., {Hariharan}, K., {Kathiravan}, C., \&
  {Gopalswamy}, N. 2016, \apj, 832, 59

\bibitem[{{Kishore} {et~al.}(2015){Kishore}, {Ramesh}, {Kathiravan}, \&
  {Rajalingam}}]{Kishore2015}
{Kishore}, P., {Ramesh}, R., {Kathiravan}, C., \& {Rajalingam}, M. 2015,
  \solphys, 290, 2409

\bibitem[{{Kumari} {et~al.}(2017a){Kumari}, {Ramesh}, {Kathiravan}, \&
  {Gopalswamy}}]{Anshu2017a}
{Kumari}, A., {Ramesh}, R., {Kathiravan}, C., \& {Gopalswamy}, N. 2017a, \apj,
  843, 10

\bibitem[{{Kumari} {et~al.}(2017b){Kumari}, {Ramesh}, {Kathiravan}, \&
  {Wang}}]{Anshu2017b}
{Kumari}, A., {Ramesh}, R., {Kathiravan}, C., \& {Wang}, T.~J. 2017b, \solphys,
  292, 161

\bibitem[{{Kumari} {et~al.}(2017c){Kumari}, {Ramesh}, {Kathiravan}, \&
  {Wang}}]{Anshu2017c}
---. 2017c, \solphys, 292, 177

\bibitem[{{Kwon} {et~al.}(2013){Kwon}, {Kramar}, {Wang}, {Ofman}, {Davila},
  {Chae}, \& {Zhang}}]{Kwon2013b}
{Kwon}, R.-Y., {Kramar}, M., {Wang}, T.~J., {et~al.} 2013, \apj, 776, 55

\bibitem[{{Lemen} {et~al.}(2012){Lemen}, {Title}, {Akin}, {Boerner}, {Chou},
  {Drake}, {Duncan}, \& {Edwards}}]{Lemen2012}
{Lemen}, J.~R., {Title}, A.~M., {Akin}, D.~J., {et~al.} 2012, \solphys, 275, 17

\bibitem[{{Lin} {et~al.}(2004){Lin}, {Kuhn}, \& {Coulter}}]{Lin2004}
{Lin}, H., {Kuhn}, J.~R., \& {Coulter}, R. 2004, \apj, 613, L177

\bibitem[{{Lin} {et~al.}(2000){Lin}, {Penn}, \& {Tomczyk}}]{Lin2000}
{Lin}, H., {Penn}, M.~J., \& {Tomczyk}, S. 2000, \apjl, 541, L83

\bibitem[{{Ma} {et~al.}(2011){Ma}, {Raymond}, {Golub}, {Lin}, {Chen}, {Grigis},
  {Testa}, \& {Long}}]{Ma2011}
{Ma}, S., {Raymond}, J.~C., {Golub}, L., {et~al.} 2011, \apj, 738, 160

\bibitem[{Mancuso {et~al.}(2019)Mancuso, Frassati, Bemporad, \&
  Barghini}]{Mancuso2019}
Mancuso, S., Frassati, F., Bemporad, A., \& Barghini, D. 2019, \aap, 624, L2

\bibitem[{{Mancuso} \& {Garzelli}(2013)}]{Mancuso2013_1}
{Mancuso}, S., \& {Garzelli}, M.~V. 2013, \aap, 560, L1

\bibitem[{{Mancuso} {et~al.}(2003){Mancuso}, {Raymond}, {Kohl}, {Ko}, {Uzzo},
  \& {Wu}}]{Mancuso2003}
{Mancuso}, S., {Raymond}, J.~C., {Kohl}, J., {et~al.} 2003, \aap, 400, 347

\bibitem[{{Mann} {et~al.}(1995){Mann}, {Classen}, \& {Aurass}}]{Mann1995}
{Mann}, G., {Classen}, T., \& {Aurass}, H. 1995, \aap, 295, 775

\bibitem[{Mann {et~al.}(2003)Mann, Klassen, Aurass, \&
  Classen}]{mann2003formation}
Mann, G., Klassen, A., Aurass, H., \& Classen, H.-T. 2003, Astronomy \&
  Astrophysics, 400, 329

\bibitem[{{Mercier} {et~al.}(2015){Mercier}, {Subramanian}, {Chambe}, \&
  {Janardhan}}]{Mercier2015}
{Mercier}, C., {Subramanian}, P., {Chambe}, G., \& {Janardhan}, P. 2015, \aap,
  576, 136

\bibitem[{{Monstein} {et~al.}(2007){Monstein}, {Ramesh}, \&
  {Kathiravan}}]{Monstein2007}
{Monstein}, C., {Ramesh}, R., \& {Kathiravan}, C. 2007, Bull. Astron. Soc.
  India, 35, 473

\bibitem[{{Mugundhan} {et~al.}(2017){Mugundhan}, {Hariharan}, \&
  {Ramesh}}]{Mugundhan2017}
{Mugundhan}, V., {Hariharan}, K., \& {Ramesh}, R. 2017, \solphys, 292, 155

\bibitem[{{Mugundhan} {et~al.}(2016){Mugundhan}, {Ramesh}, {Indrajit V.
  Barve}., {Kathiravan}, {Gireesh}, {Kharb}, \& {Misra}}]{Mugundhan2016}
{Mugundhan}, V., {Ramesh}, R., {Indrajit V. Barve}., {et~al.} 2016, \apj, 831,
  154

\bibitem[{{Mugundhan} {et~al.}(2018){Mugundhan}, {Ramesh}, {Kathiravan},
  {Gireesh}, {Kumari}, {Hariharan}, \& {Indrajit V. Barve}.}]{Mugundhan2018}
{Mugundhan}, V., {Ramesh}, R., {Kathiravan}, C., {et~al.} 2018, \apjl, 855, L8

\bibitem[{{Patsourakos} \& {Vourlidas}(2012)}]{Patsourakos2012}
{Patsourakos}, S., \& {Vourlidas}, A. 2012, \solphys, 281, 187

\bibitem[{{Pomoell} {et~al.}(2009){Pomoell}, {Vainio}, \&
  {Pohjolainen}}]{Pomoell2009}
{Pomoell}, J., {Vainio}, R., \& {Pohjolainen}, S. 2009, in Universal
  Heliophysical Processes, ed. N.~{Gopalswamy} \& D.~F. {Webb}, Proc. IAU Symp.
  257, 493

\bibitem[{{Ramesh}(2011)}]{Ramesh2011a}
{Ramesh}, R. 2011, in Proc. of the 1st Asia-Pacific Sol. Phys. Meeting, ed.
  A.~R. {Choudhuri} \& D.~{Banerjee}, Astron. Soc. India Conf. Ser. 2, 55

\bibitem[{{Ramesh} {et~al.}(2012a){Ramesh}, {Anna Lakshmi}, {Kathiravan},
  {Gopalswamy}, \& {Umapathy}}]{Ramesh2012a}
{Ramesh}, R., {Anna Lakshmi}, M., {Kathiravan}, C., {Gopalswamy}, N., \&
  {Umapathy}, S. 2012a, \apj, 752, 107

\bibitem[{{Ramesh} {et~al.}(2012b){Ramesh}, {Kathiravan}, {Indrajit V. Barve}.,
  \& {Rajalingam}}]{Ramesh2012b}
{Ramesh}, R., {Kathiravan}, C., {Indrajit V. Barve}., \& {Rajalingam}, M.
  2012b, \apj, 744, 165

\bibitem[{{Ramesh} {et~al.}(2010a){Ramesh}, {Kathiravan}, \&
  {Sastry}}]{Ramesh2010a}
{Ramesh}, R., {Kathiravan}, C., \& {Sastry}, C.~V. 2010a, \apj, 711, 1029

\bibitem[{{Ramesh} {et~al.}(2004){Ramesh}, {Kathiravan}, \& {Satya
  Narayanan}}]{Ramesh2004}
{Ramesh}, R., {Kathiravan}, C., \& {Satya Narayanan}, A. 2004, Asian J. Phys.,
  13, 277

\bibitem[{{Ramesh} {et~al.}(2011b){Ramesh}, {Kathiravan}, \& {Satya
  Narayanan}}]{Ramesh2011b}
---. 2011b, \apj, 734, 39

\bibitem[{{Ramesh} {et~al.}(2003){Ramesh}, {Kathiravan}, {Satya Narayanan}, \&
  {Ebenezer}}]{Ramesh2003}
{Ramesh}, R., {Kathiravan}, C., {Satya Narayanan}, A., \& {Ebenezer}, E. 2003,
  \aap, 400, 753

\bibitem[{{Ramesh} {et~al.}(2010b){Ramesh}, {Kathiravan}, {Sreeja S. Kartha},
  \& {Gopalswamy}}]{Ramesh2010b}
{Ramesh}, R., {Kathiravan}, C., {Sreeja S. Kartha}, \& {Gopalswamy}, N. 2010b,
  \apj, 712, 188

\bibitem[{{Ramesh} {et~al.}(2008){Ramesh}, {Kathiravan}, {Sundara Rajan},
  {Indrajit V. Barve}, \& {Sastry}}]{Ramesh2008}
{Ramesh}, R., {Kathiravan}, C., {Sundara Rajan}, M.~S., {Indrajit V. Barve}, \&
  {Sastry}, C.~V. 2008, \solphys, 253, 319

\bibitem[{{Ramesh} {et~al.}(2013){Ramesh}, {Kishore}, {Sargam M. Mulay},
  {Barve}, {Kathiravan}, \& {Wang}}]{Ramesh2013}
{Ramesh}, R., {Kishore}, P., {Sargam M. Mulay}, {et~al.} 2013, \apj, 778, 30

\bibitem[{{Ramesh} {et~al.}(2006a){Ramesh}, {Nataraj}, {Kathiravan}, \&
  {Sastry}}]{Ramesh2006a}
{Ramesh}, R., {Nataraj}, H.~S., {Kathiravan}, C., \& {Sastry}, C.~V. 2006a,
  \apj, 648, 707

\bibitem[{Ramesh {et~al.}(1998)Ramesh, Subramanian, Sundara~Rajan, \&
  Sastry}]{Ramesh1998}
Ramesh, R., Subramanian, K., Sundara~Rajan, M., \& Sastry, C.~V. 1998,
  \solphys, 181, 439

\bibitem[{{Ramesh} {et~al.}(1999a){Ramesh}, {Subramanian}, \&
  {Sastry}}]{Ramesh1999a}
{Ramesh}, R., {Subramanian}, K.~R., \& {Sastry}, C.~V. 1999a, \aaps, 139, 179

\bibitem[{{Ramesh} {et~al.}(1999b){Ramesh}, {Subramanian}, \&
  {Sastry}}]{Ramesh1999b}
---. 1999b, \solphys, 185, 77

\bibitem[{{Ramesh} {et~al.}(2006b){Ramesh}, {Sundara Rajan}, \&
  {Sastry}}]{Ramesh2006b}
{Ramesh}, R., {Sundara Rajan}, M.~S., \& {Sastry}, C.~V. 2006b, Exp. Astron.,
  21, 31

\bibitem[{{Sasikumar Raja} {et~al.}(2016){Sasikumar Raja}, {Ingale}, {Ramesh},
  {Subramanian}, {Manoharan}, \& {Janardhan}}]{Sasikumar2016}
{Sasikumar Raja}, K., {Ingale}, M., {Ramesh}, R., {et~al.} 2016, J. Geophys.
  Res.: Space Phys., 121, 11605

\bibitem[{{Sasikumar Raja} {et~al.}(2013a){Sasikumar Raja}, {Kathiravan},
  {Ramesh}, {Rajalingam}, \& {Indrajit V. Barve}}]{Sasi2013a}
{Sasikumar Raja}, K., {Kathiravan}, C., {Ramesh}, R., {Rajalingam}, M., \&
  {Indrajit V. Barve}. 2013a, \apjs, 207, 2

\bibitem[{{Sasikumar Raja} \& {Ramesh}(2013b)}]{Sasikumar2013b}
{Sasikumar Raja}, K., \& {Ramesh}, R. 2013b, \apj, 775, 38

\bibitem[{{Sasikumar Raja} {et~al.}(2014){Sasikumar Raja}, {Ramesh},
  {Hariharan}, {Kathiravan}, \& {Wang}}]{Sasikumar2014}
{Sasikumar Raja}, K., {Ramesh}, R., {Hariharan}, K., {Kathiravan}, C., \&
  {Wang}, T.~J. 2014, \apj, 796, 56

\bibitem[{{Sastry}(2009)}]{Sastry2009}
{Sastry}, C.~V. 2009, \apj, 697, 1934

\bibitem[{{Smerd} {et~al.}(1974){Smerd}, {Sheridan}, \& {Stewart}}]{Smerd1974}
{Smerd}, S.~F., {Sheridan}, K.~V., \& {Stewart}, R.~T. 1974, in Coronal
  Disturbances, ed. G.~A. {Newkirk}, Proc. IAU Symp. 57, 389

\bibitem[{{Smerd} {et~al.}(1975){Smerd}, {Sheridan}, \& {Stewart}}]{Smerd1975}
{Smerd}, S.~F., {Sheridan}, K.~V., \& {Stewart}, R.~T. 1975, Astrophys. Lett.,
  16, 23

\bibitem[{{Stewart} \& {McLean}(1982)}]{Stewart1982}
{Stewart}, R.~T., \& {McLean}, D.~J. 1982, Pub. Astron. Soc. Aust., 4, 386

\bibitem[{{Tomczyk} {et~al.}(2008){Tomczyk}, {Card}, {Darnell}, {Elmore},
  {Lull}, {Nelson}, {Streander}, {Burkepile}, {Casini}, \&
  {Judge}}]{Tomczyk2008}
{Tomczyk}, S., {Card}, G.~L., {Darnell}, T., {et~al.} 2008, \solphys, 247, 411

\bibitem[{{Tun} \& {Vourlidas}(2013)}]{Tun2013}
{Tun}, S.~D., \& {Vourlidas}, A. 2013, \apj, 766, 130

\bibitem[{{Veronig} {et~al.}(2010){Veronig}, {Muhr}, {Kienreich}, {Temmer}, \&
  {Vr{\v s}nak}}]{veronig2010}
{Veronig}, A.~M., {Muhr}, N., {Kienreich}, I.~W., {Temmer}, M., \& {Vr{\v
  s}nak}, B. 2010, \apjl, 716, L57

\bibitem[{{Vr{\v s}nak} \& {Cliver}(2008)}]{Vrsnak2008}
{Vr{\v s}nak}, B., \& {Cliver}, E.~W. 2008, \solphys, 253, 215

\bibitem[{{Vr{\v s}nak} {et~al.}(2002){Vr{\v s}nak}, {Magdaleni{\'c}},
  {Aurass}, \& {Mann}}]{Vrsnak2002}
{Vr{\v s}nak}, B., {Magdaleni{\'c}}, J., {Aurass}, H., \& {Mann}, G. 2002,
  \aap, 396, 673

\bibitem[{{Wang} \& {Davila}(2014)}]{Wang2014}
{Wang}, T., \& {Davila}, J.~M. 2014, \solphys, 289, 3723

\bibitem[{{Wang} {et~al.}(2017){Wang}, {Reginald}, {Davila}, {St.~Cyr}, \&
  {Thompson}}]{Wang2017}
{Wang}, T., {Reginald}, N.~L., {Davila}, J.~M., {St.~Cyr}, O.~C., \&
  {Thompson}, W.~T. 2017, \solphys, 292, 97

\bibitem[{Warmuth \& Mann(2005)}]{warmuth2005model}
Warmuth, A., \& Mann, G. 2005, Astronomy \& Astrophysics, 435, 1123

\bibitem[{{Wiegelmann} {et~al.}(2017){Wiegelmann}, {Petrie}, \&
  {Riley}}]{Wiegelmann2017}
{Wiegelmann}, T., {Petrie}, G.~J.~D., \& {Riley}, P. 2017, \ssr, 210, 249

\bibitem[{{Zhang} {et~al.}(2012){Zhang}, {Cheng}, \& {Ding}}]{Zhang2012}
{Zhang}, J., {Cheng}, X., \& {Ding}, M. 2012, Nat. Comm., 3, 747

\bibitem[{{Zimovets} {et~al.}(2012){Zimovets}, {Vilmer}, {Chian}, {Sharykin},
  \& {Struminsky}}]{Zimovets2012}
{Zimovets}, I., {Vilmer}, N., {Chian}, A.~C.-L., {Sharykin}, I., \&
  {Struminsky}, A. 2012, \aap, 547, A6

\bibitem[{{Zucca} {et~al.}(2014{\natexlab{a}}){Zucca}, {Carley}, {Bloomfield},
  \& {Gallagher}}]{Zucca2014_1}
{Zucca}, P., {Carley}, E.~P., {Bloomfield}, D.~S., \& {Gallagher}, P.~T.
  2014{\natexlab{a}}, \aap, 564, A47

\bibitem[{{Zucca} {et~al.}(2014{\natexlab{b}}){Zucca}, {Pick}, {Demoulin},
  {Kerdraon}, {Lecacheux}, \& {Gallagher}}]{Zucca2014}
{Zucca}, P., {Pick}, M., {Demoulin}, P., {et~al.} 2014{\natexlab{b}}, \apj,
  795, 68

\end{thebibliography}

\end{document}